\documentstyle[prl,aps,twocolumn,epsfig]{revtex}

\def\be{\begin{equation}}
\def\ee{\end{equation}}
\def\bea{\begin{eqnarray}}
\def\eea{\end{eqnarray}}
\def\ba{\begin{array}}
\def\ea{\end{array}}

\def\0{$\Gamma_0$}

\def\p{\phi}

\def\vt{\vartheta}

\def\Mb{M\"obius }

\begin{document}

\twocolumn[\hsize\textwidth\columnwidth\hsize\csname @twocolumnfalse\endcsname

\title{Ising model on nonorientable surfaces: \\
Exact solution for the M\"obius strip and the Klein bottle}
\author{Wentao T. Lu and F. Y. Wu}
\address{Department of Physics\\
 Northeastern University, Boston,
Massachusetts 02115}

\date{\today}

\maketitle

\begin{abstract}
Closed-form expressions are obtained for the partition function
of the Ising model on an ${\cal M}\times {\cal N}$  simple-quartic 
lattice embedded on a M\"obius strip and a Klein bottle
for finite ${\cal M}$ and ${\cal N}$.
The finite-size effects at criticality are analyzed and compared 
with those under cylindrical and toroidal boundary conditions.
Our analysis confirms that the central charge is $c=1/2$.
\end{abstract}


\vskip1.5pc]

\section{Introduction}
There has been considerable recent interest \cite{luwu99,tzeng,tesler} 
in studying lattice models on nonorientable surfaces, both as new 
challenging unsolved lattice-statistical problems and as a 
realization and testing of predictions of the conformal field theory  
\cite{cardy}.  In a recent paper \cite{luwu99} we have
presented the solution  of dimers on the \Mb strip and
Klein bottle and studied its finite-size corrections. 
In this paper we consider the Ising model.

The Ising model in two dimensions was first solved by Onsager in 1944
\cite{onsager} who obtained the close-form expression of the 
partition function for a simple-quartic ${\cal M}\times {\cal N}$ 
lattice wrapped on a  cylinder. The exact solution for an 
${\cal M}\times {\cal N}$ lattice on a torus, namely,
with periodic boundary condition in both directions, was obtained 
by Kaufman 4 years later \cite{kauf}.
Onsager and  Kaufman used spinor analysis to  derive the solutions, 
and the solution under the cylindrical boundary condition
was rederived later by McCoy and Wu \cite{mccoywu} using the 
method of dimers. As far as we know, these are the only known 
solutions of the two-dimensional Ising model on finite lattices.  
Here, using the method of dimers, we derive exact expressions 
for the partition function of the Ising model on finite \Mb strips 
and  Klein bottles. As we shall see, as a consequence of 
the \Mb topology, the solution assumes a form which depends on
whether the width of the lattice is even or odd. However,
all solutions yield the same bulk free energy.
We also present results of finite-size analyses for corrections 
to the bulk solution, and compare with those deduced under other 
boundary conditions. Our explicit calculations confirm that the 
central charge is $c=1/2$.

\section{The $2M\times N$ M\"obius strip}
To begin with, we consider  a $2M \times N$ simple-quartic Ising 
lattice  ${\cal L}$ embedded on a M\"obius strip, where $M,N$ 
are integers and $2M$ is the width of the strip. The example of 
a lattice ${\cal L}$ for $2M=4, N=5$ is shown in Fig. 1.

While we shall consider the case of a uniform reduced interaction
$K$, to facilitate considerations it is convenient to let the  $N$ 
vertical edges located in the middle  of the strip to take on a 
different interaction $K_1$ as shown. By setting $K_1=0$ the \Mb 
strip reduces to an $M\times 2N$ strip with a ``cylindrical" 
boundary condition, namely, periodic in one direction and free
in the other, for which the partition function has been evaluated 
by McCoy and Wu \cite{mccoywu}. By setting $K_1=\infty$ the two 
center rows of spins coalesce into a single row
with an (additive) interaction which in this case is $2K$.  
These are two key elements of our consideration.

\begin{figure}[htbp]
\center{\rule{5cm}{0.mm}}
\rule{5cm}{0.mm}
\vskip -.8cm
\epsfig{figure=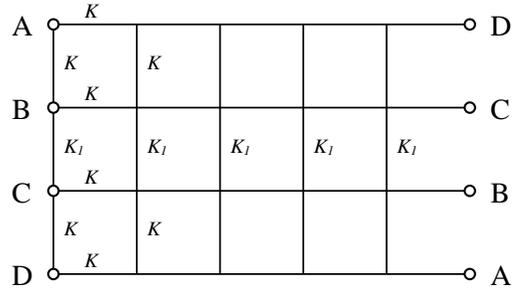,width=2.8in}
\vskip .2cm
\caption{A $4\times 5$ \Mb strip ${\cal L}$.  
Vertices labeled $A, B, C, D$  are repeated sites.}
\label{fig1}
\end{figure}

Following  standard procedures \cite{mccoywu} we write the partition 
function of the Ising model on ${\cal L}$ as
\bea
Z^{\rm Mob}_{2M,N}(K, K_1)&=&
2^{2MN}(\cosh K)^{2(2MN-N)}(\cosh K_1)^{N}\nonumber \\
&&\quad \times G(z,z_1),\label{part}
\eea
where $z=\tanh K,\ z_1=\tanh K_1$, and
\be
G(z,z_1)=\sum_{\rm closed\ polygons} z^nz_1^{n_1} \label{gen}
\ee
is the generating function of all closed polygonal graphs on 
${\cal L}$ with edge weights $z$ and $z_1$.  Here, $n$ is the 
number of polygon edges with weight $z$ and $n_1$ the number 
of edges with weight $z_1$.

The generating function $G(z, z_1)$ is a multinomial in $z$ and $z_1$ 
and, due to the \Mb topology, the integer $n_1$ can take on any value 
in $\{0,N\}$. Thus, we have
\be
G(z,z_1)=\sum_{n_1=0}^{N} T_{n_1}(z) z_1^{n_1}, \label{gen1}
\ee
where $T_{n_1}(z)$ are polynomials in $z$ with strictly positive 
coefficients.

To evaluate $G(z,z_1)$, we again follow the usual
procedure of  mapping  polygonal configurations on ${\cal L}$ onto 
dimer configurations
on a dimer lattice ${\cal L}_D$ of $8MN$ sites, constructed by expanding 
each site of ${\cal L}$  into a  ``city" of 4
sites \cite{mccoywu,kast}.  The resulting ${\cal L}_D$ for
the $4\times 5$ ${\cal L}$ is
 shown in Fig. 2. 

\begin{figure}[htbp]
\center{\rule{5cm}{0.mm}}
\rule{5cm}{0.mm}
\vskip -1.2cm
\epsfig{figure=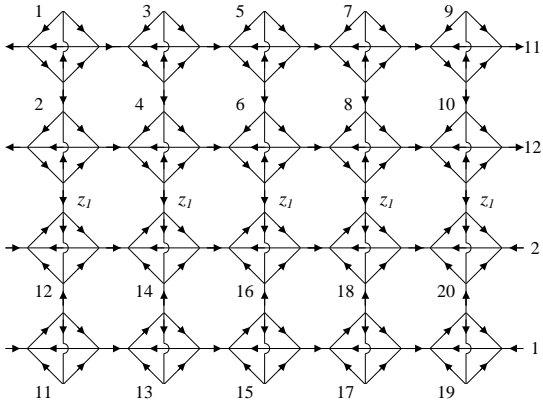,width=3.2in}
\vskip -1.6in
\caption{The dimer lattice ${\cal L}_D$ corresponding to 
the $4\times 5$ \Mb strip.}
\label{fig2}
\end{figure}

As  the deletion of all $z_1$ edges reduces 
the lattice to one with a cylindrical boundary
condition  solved in \cite{mccoywu},
we orient all edges of weights $z$ and 1
as in \cite{mccoywu}.  In addition,  all $z_1$ edges are oriented in the
direction as shown in Fig. 2.
Then, we have the following:

\bigskip
\noindent
{\it Theorem:

   Let $A$ be the $8MN\times 8MN$ antisymmetric determinant defined by
the lattice edge orientation  shown in Fig. 2, and let 
\be
 {\rm Pf} A (z,z_1) = 
\sqrt {{\rm det} A(z,z_1)} \label{pfaffian}
\ee
denote the Pfaffian of  $A$.  Then 
\be
{\rm Pf} A (z,z_1) = \sum_{n_1=0}^{N} \epsilon_{n_1}
T_{n_1}(z) z_1^{n_1}, \label{theorem}
\ee
where 
$\epsilon_{4m}=\epsilon_{4m+1}=1,\quad \epsilon_{4m+2}=\epsilon_{4m+3}=-1$ for
any integer $m\geq 0$.}

\bigskip
\noindent
Remark:
Define
\be
X_p= \sum_{m=0}^{[N/4]} T_{4m+p}(z) z_1^{4m+p}, \quad p=0,1,2,3,
\ee
where $[N/4]$ is the integral part of $N/4$
so that $G(z,z_1) =X_0+X_1+X_2+X_3$.  
It then follows from (\ref{theorem}) that we have
\be
{\rm Pf}A(z,\pm iz_1) = X_0 + X_2 \pm i( X_1 +  X_3).
\ee
As a consequence, we obtain 
\bea
G(z,z_1)&=&{1\over 2}\Big[(1-i){\rm Pf}A(z,iz_1)\nonumber \\
&&\quad +(1+i)
{\rm Pf}A(z,-iz_1)\Big],\label{gen2}
\eea
where,  as evaluated in the
next section,  the Pfaffian is given by
\bea
&&{\rm Pf}A(z,z_1)=[z(1-z^2)]^{MN}\nonumber \\
&&\quad \times \prod_{n=1}^{N}
\Bigg[{\sinh(M+1)t(\phi_n)-c(z,z_1)\sinh Mt(\phi_n)\over 
\sinh t(\phi_n)}\Bigg], 
\label{faf}
\eea
with
\bea
c(z,z_1)&=&{z(1+z^2+2z\cos\p_n)+2(-1)^nz_1\sin\p_n \over 1-z^2}\nonumber \\
\cosh t(\p)&=& \cosh 2K \coth 2K - \cos \p  \nonumber \\
 \p_n&=& (2n-1)\pi/2N.
\eea
Here we have used the fact that $\prod_{n=1}^{N}=\prod_{n=N+1}^{2N}$
in the product in (\ref{faf}).
Substituting these results into (\ref{part}),
we are led to the following explicit expression for the 
partition function,
\bea
Z^{\rm Mob}_{2M,N}(K, K)&=&
{1\over 2}\Big(2\sinh 2K\Big)^{MN} \nonumber \\
&&\quad \times\Big[(1-i) F_+ + (1+i) F_- \Big] ,\label{partmob}
\eea
where
\bea 
F_\pm &=& \prod_{n=1}^N \Bigg[ e^{Mt(\p_n)}
\Bigg({{e^{t(\p_n)}- c(z,\pm iz)}\over {2\sinh t(\p_n)}}\Bigg)\nonumber \\
&&\qquad-e^{-Mt(\p_n)}\Bigg({{e^{-t(\p_n)}- c(z,\pm iz)}\over {2\sinh t(\p_n)}}\Bigg) \Bigg].  \label{fpm}
\eea
 This completes the evaluation of the Ising partition function for the $2M\times N$ \Mb
strip.  Note that we have $\cosh t(\p_n) \geq 1$ so we can always take $t(\p_n) \geq 0$.
The leading contribution  in (\ref{fpm}) for large $M$ is therefore
$\prod e^{Mt(\p_n)}$. 

For the $2\times 5$ \Mb strip, for example, we find
\bea
{\rm Pf} A(z,z_1) &=& 1+z^{10}+10z_1z^5-5z_1^2z^2(1+z^2+z^4+z^6)\nonumber \\
&&-20z_1^3z^5+5z_1^4z^4(1+z^2)+2z_1^5z^5,  \nonumber \\
        G(z,z_1)  &=&1+z^{10}+10z_1z^5+5z_1^2z^2(1+z^2+z^4+z^6)\nonumber \\
&&+20z_1^3z^5+5z_1^4z^4(1+z^2)+2z_1^5z^5,
\eea
which can be verified by explicit enumerations.

We next prove  the theorem.
 
Considered as a multinomial in $z$
and $z_1$, there exists
a one-one correspondence  between terms in  the dimer generating function $G(z,z_1)$
and (combinations of) terms in the Pfaffian
(\ref{pfaffian}).
However,  while all terms in $G(z,z_1)$ are positive,
terms in the Pfaffian do not necessarily
possess the same sign.  The crux of the matter is to find 
an appropriate linear combination of 
  Pfaffians  to yield the desired $G(z,z_1)$.  For this purpose
it is  convenient to 
compare an arbitrary term $C_1$ in the Pfaffian with a standard one $C_0$. 
 We choose  $C_0$ to be one
in which  no $z$ and $z_1$ dimers  are present.

The superposition of  two dimer configurations represented 
by $C_0$ and $C_1$ produces
superposition polygons. Kasteleyn \cite{kasteleyn} has shown that the
 two  terms
will have the same sign if all superposition polygons 
are oriented ``clockwise-odd", namely, there is an odd number of edges oriented
in the clockwise direction. 

Now since all $z$ and $1$ edges of ${\cal L}_D$  are oriented as
 in \cite{mccoywu},  terms in the Pfaffian with no $z_1$ edges ($n_1=0$) 
will have the same sign as $C_0$.  
 To determine the  sign of a term when $z_1$ edges are present, 
we associate a $+$ sign to each
clockwise-odd superposition polygon and a $-$ sign to each
clockwise-even superposition polygon. 
 Then the sign of $C_1$ relative to $C_0$ is the product of the signs of all
superposition polygons.  The
following elementary facts   can be readily verified:

i) Deformations of the borders of a superposition polygon always change
$m_1$, the number of {\it its} $z_1$ edges, by multiples of 2.

ii) The sign of a superposition polygon is reversed under border deformations
   which  change $m_1$   by 2.
  
iii) Superposition polygons having 0 or 1 $z_1$ edges have a sign $+$.

iv) There can be at most one superposition polygon 
having an odd number 
of $z_1$ edges (a property unique to nonorientable surfaces).

Let $m_1=4m+p$, where 
$m$ is an integer and $p=0,1,2,3$. 
Because of iv), we need only to consider the presence of
at most one polygon having
$p=1$ or $3$.
 It now follows from i)  and iii) 
that $\epsilon_{4m}=\epsilon_{4m+1}=+$, and from i), ii) and iii) 
that $\epsilon_{4m+2}=\epsilon_{4m+3}=-$.  This establishes the theorem.

\section{Evaluation of the Pfaffian}
In this section we derive the expression (\ref{faf}).

From the edge orientation of  ${\cal L}_D$ 
of Fig. 2, one finds that the $8MN \times 8MN$ antisymmetric
matrix $A$ assumes the form
\bea
A(z,z_1)&=&A_0(z)\otimes I_{2N}+A_+(z)\otimes J_{2N}
+A_-(z)\otimes J^T_{2N}\nonumber \\
&&\quad +A_1(z_1)\otimes H_{2N},  \label{mobpf}
\eea
where $A_0, A_+, A_-, A_1$ are 
 $4M\times 4M$ matrices, $I_{2N}$ is the $2N\times 2N$ identity
matrix, and $ J_{2N}, H_{2N}$ are the $2N\times 2N$ matrices
\bea
J_{2N} &=& \pmatrix {0 & 1 & 0 &\cdots & 0 \cr
                  0 & 0  & 1 &\cdots & 0 \cr
                  \vdots & \vdots & \vdots & \ddots & \vdots \cr
                  0 & 0  & 0 &\cdots & 1 \cr
-1 & 0  & 0 &\cdots & 0},\nonumber \\
H_{2N} &=& \pmatrix{0&I_N\cr -I_N&0}.
\eea
In addition, one has
\bea
A_0(z)&=&a_{0,0}\otimes I_M+a_{0,1}(z)\otimes F_M+a_{0,-1}(z)\otimes F^T_M\nonumber \\
A_\pm(z)&=&a_{\pm 1,0}(z)\otimes I_M \nonumber \\
A_1(z_1) &=& a(z_1)\otimes G_M,
\eea
where  $F_M, G_M$ are $M\times M$  matrices
\bea
F_M&=& \pmatrix {0 & 1 & 0 &\cdots & 0 \cr
                  0 & 0  & 1 &\cdots & 0 \cr
                  \vdots & \vdots & \vdots & \ddots & \vdots \cr
                  0 & 0  & 0 &\cdots & 1 \cr
0 & 0  & 0 &\cdots & 0},\nonumber \\
G_M &=& \pmatrix {0 & 0  &\cdots & 0 & 0 \cr
                  0 & 0  &\cdots & 0& 0 \cr
                  \vdots  & \vdots & \ddots & \vdots& \vdots \cr
                  0 & 0  &\cdots & 0 & 0 \cr
0 & 0   &\cdots & 0 & 1},
\eea
$F^T_M$ is the transpose of $F_M$,
and
\bea
a_{0,0}&=&\pmatrix{0&1&-1&-1\cr
-1&0&1&-1\cr
1&-1&0&1\cr
1&1&-1&0\cr},\nonumber \\
a(z_1)&=&\pmatrix{0&0&0&0\cr0&0&0&0\cr 0&0&z_1&0\cr 0&0&0&0\cr}
\nonumber  \\
a_{1,0}(z)&=&\pmatrix{0&z&0&0\cr 0&0&0&0\cr 0&0&0&0\cr0&0&0&0\cr},
\nonumber \\
a_{0,1}(z)&=&\pmatrix{0&0&0&0\cr 0&0&0&0\cr 0&0&0&z\cr0&0&0&0\cr}, 
\nonumber \\
a_{-1,0}(z)&=&a^T_{1,0}(z) \nonumber  \\
a_{0,-1}(z)&=&a^T_{0,1}(z).
\eea

We use the fact that the determinant in (\ref{pfaffian})
is equal to  the product
of the eigenvalues of the matrix $A$.
To evaluate the latter, we note that $J_{2N},J^T_{2N}$ and $H_{2N}$ mutually commute so that
they can be diagonalized simultaneously.  This leads to the respective eigenvalues
$e^{i\p_n}, e^{-i\p_n}$ and $i(-1)^{n+1}$ and the expression 
\be
{\rm det}A(z,z_1)=\prod_{n=1}^{2N} {\rm det}A_M(z, z_1;\p_n), \label{det}
\ee
where 
\bea
 A_M(z,z_1;\p_n)&=&A_0(z)+A_+(z)e^{i\p_n}+A_-(z)e^{-i\p_n}\nonumber \\
&&\quad + i(-1)^{n+1}A_1(z_1)
\eea
is a $4M \times 4M$ matrix.
Writing out explicitly, we have
\bea
&& A_{M}(z,z_1;\p_n)= \nonumber \\
&&\pmatrix{ B(z) &   a_{0,1}(z)    & 0& & \cr
                    a_{0,-1}(z) & B(z) & a_{0,1}(z)&0 & \cr
                       \vdots      &     \vdots& \ddots&\vdots &\vdots \cr
                                &      0  &a_{0,-1}(z) & B(z) & a_{0,1}(z) \cr
                                   &        &    0  & a_{0,-1}(z)& C(z,z_1) \cr} 
\label{am}\nonumber \\
\eea
where $C(z,z_1)=B(z)+i(-1)^{n+1}a(z_1)$, and
\be
B(z)= \pmatrix{0&1+ze^{i\p_n}&-1&-1 \cr -(1+ze^{-i\p_n})&0&1&-1 \cr 
1&-1&0& 1 \cr                1&1& -1&0 }.
\ee
 
The evaluation of ${\rm det}A_{M}(z,z_1;\p_n)$  can be carried out by using a
recursion procedures introduced in
 \cite{luwuising} for a self-dual Ising model. 
Specifically, let  $B_{M}={\rm det}A_{M}(z,z_1;\p_n)$ and $D_M$ 
the determinant of the matrix $A_M(z,z_1;\p_n)$ with the fourth row and fourth column removed.
Then by expanding the determinants one finds  the  recursion relation 
(which is the same as that in \cite{luwuising})
\be
\pmatrix{B_{M}\cr D_M\cr}=
  \pmatrix {a_{11} & a_{12}\cr a_{21} & a_{22} \cr}   
  \pmatrix{B_{M-1}\cr D_{M-1}\cr}, \quad  M\geq 2, \label{rec}
\ee
with 
\bea
a_{11}&=&1+z^2-2z\cos\p_n,\nonumber \\ 
a_{12}&=&-2iz^3\sin\p_n,\nonumber \\
a_{21}&=&2iz\sin\p_n,\nonumber \\
a_{22}&=&z^2(1+z^2+2z\cos\p_n),
\eea
and the initial condition (which is different from \cite{luwuising})
\bea
B_{1}&=& B_1(z,z_1)\nonumber \\ 
&\equiv& 1-2z\cos \p_n +z^{2}-2(-1)^nzz_1\sin\p_n,\nonumber \\ 
D_{1}&=& D_1(z,z_1) \nonumber \\
&\equiv& 2iz\sin \p_n - i(-1)^nz_1(1+2z\cos \p_n +z^{2}). \label{initial}
\eea
This leads to the solution
 \bea
B_{M}&=&B_1{\frac{\lambda_+^{M}-\lambda_-^{M}}{\lambda_+-\lambda_-}}
-(a_{22}B_1-a_{12}D_1){\frac{\lambda_+^{M-1}-\lambda_-^{M-1}}{
\lambda_+-\lambda_-}},\nonumber \\
D_{M}&=&D_1{\frac{\lambda_+^{M}-\lambda_-^{M}}{\lambda_+-\lambda_-}}
-(a_{11}D_1-a_{21}B_1){\frac{\lambda_+^{M-1}-\lambda_-^{M-1}}{
\lambda_+-\lambda_-}}, \label{solution}\nonumber \\
\eea
 where $\lambda_\pm=z(1-z^2)e^{\pm t(\p_n)}$ are the eigenvalues of the $2\times 2$ matrix in 
(\ref{rec}).
 After some algebraic manipulation, this yields the expression (\ref{faf})
quoted in the preceding section.

\section{The $(2M-1)\times N$ M\"obius strip}
 We consider a $(2M-1)\times N$  \Mb strip in this section.

In order to make use of results of the preceding sections, we 
start from the $2M \times N$ strip of Sec. 2, and let  spins in 
the two center rows of the
strip (the $M$th and $(M+1)$th rows) having interactions  $K_0=K/2$.
The example of a $4\times 5$ lattice with these interactions is shown in Fig. 3. 
Then, by taking $K_1=\infty$ ($z_1=1$) as described in Sec. 1, 
this lattice reduces to the desired
 $(2M-1)\times N$ \Mb strip of a uniform interaction $K$.

\begin{figure}[htbp]
\center{\rule{5cm}{0.mm}}
\rule{5cm}{0.mm}
\vskip -.8cm
\epsfig{figure=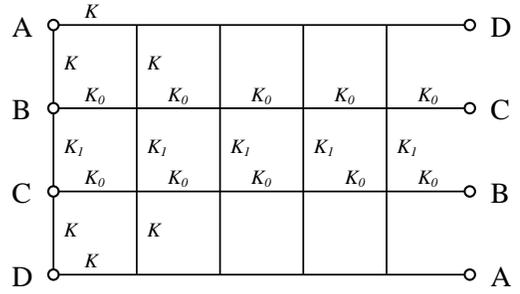,width=2.8in}
\vskip .2cm
\caption{Labelings of a $4\times 5$ \Mb strip which reduces to 
a $3\times 5$ \Mb strip upon taking $K_1=\infty$.}
\label{fig3}
\end{figure}

Following this procedure, we have
\bea
Z^{\rm Mob}_{2M-1,N}(K)&=&2^{(2M-1)N}(\cosh K)^{4(M-1)N}\nonumber \\
&&\quad \times \cosh^{2N} (K/2)G(z,z_0,z_1)\Big|_{z_1=1},\label{part1}
\eea
where $z_0=\tanh(K/2)$, and $G(z,z_0,z_1)$ is the generating function
of closed polygons on the $2M \times N$ \Mb net with edge weights
as described in the above.
   
The generating function $G(z,z_0,z_1)$ can 
be evaluated as in the previous sections. 
In place of (\ref{pfaffian}), (\ref{det}) and (\ref{am}),
we now have
\bea
{\rm Pf} A(z,z_0,z_1)&=& \sqrt {{\rm det}  A(z,z_0,z_1)}  \nonumber \\
   &=& \prod _{n=1}^{2N}\sqrt{{\rm det} A_M(z,z_0,z_1;\p_n)  } \label{pf2}
\eea
with
\bea
&& A_{M}(z,z_0,z_1;\p_n)= \nonumber \\
&&\pmatrix{ B(z) &   a_{0,1}(z)    &0 & & \cr
                    a_{0,-1}(z) & B(z) & a_{0,1}(z)&0 & \cr
                       \vdots      &     \vdots& \ddots&\vdots &\vdots \cr
                                &     0   &a_{0,-1}(z) & B(z) & a_{0,1}(z) \cr
                                   &        &   0   & a_{0,-1}(z)& C(z_0,z_1) \cr}. \nonumber \\
\label{am1}
\eea
Then (\ref{gen2}) becomes
 \bea
G(z,z_0,z_1)&=&{1\over 2}\Big[\big(1-i\big){\rm Pf}A(z,z_0,iz_1)\nonumber \\
&&\quad +\big(1+i\big)
{\rm Pf}A(z,z_0,-iz_1)\Big].\label{oddgen2}
\eea
The evaluation of 
  ${\rm det} A_M(z,z_0,z_1;\p_n)$ can again be done recursively.
  Define as before $B_M
= {\rm det} A_M(z,z_0,z_1;\p_n)$ and  $D_M$
 the determinant of  $A_M$ with the fourth
row and column removed,  one  obtains again the  recursion relations
(\ref{rec}) and arrives at precisely the same solution (\ref{solution}),  but now with 
a different initial condition
\be
B_1 = B_1(z_0, z_1), \hskip 1cm D_1 = D_1(z_0, z_1),
\ee
where the functions $B_1$ and $D_1$ are  defined in (\ref{initial}). 
 After some algebra, this leads to 
\bea
&&{\rm Pf}A(z,z_0,z_1)=[z(1-z^2)]^{(M-1)N}\nonumber \\
&&\hskip 0.5cm \times \prod_{n=1}^{N}
\Bigg[{c_1\sinh Mt(\p_n)-c_2\sinh (M-1)t(\p_n)
\over \sinh t(\p_n)}\Bigg], \label{pfa}
\eea
where
\bea
c_1&=&{2z_0\over z}\Bigg\{
1-z\Big[\cos \p_n +(-1)^nz_1\sin\p_n\Big]\Bigg\},\nonumber \\
c_2&=& 2z_0\Bigg\{1+ z\Big[\cos\p_n+(-1)^nz_1\sin\p_n\Big]\Bigg\}. 
\label{cc}
\eea
The substitution of (\ref{pfa}) into (\ref{oddgen2}) and (\ref{part1})
now  completes the evaluation of the partition function for a $(2M-1)\times N$ \Mb
strip.

\section{The Klein bottle}
The Ising model on a Klein bottle can be considered similarly.
We consider first a $2M\times N$ lattice ${\cal L}$,  constructed
by connecting the upper and lower edges of the \Mb strip of Fig. 1 in a
periodic fashion with $N$ extra vertical edges.
As in the case of the \Mb strip, it is convenient to let the extra edges 
have interactions $K_2$.  The solution for a uniform interaction $K$ is 
obtained at the end by setting $K_1=K_2=K$.

The  Ising partition function for the Klein bottle now assumes the form 
\bea
&&Z^{\rm Kln}_{2M,N}(K,K_1,K_2)=2^{2MN}(\cosh K)^{4MN-2N}\nonumber \\
&&\hskip 1.8cm \times (\cosh K_1\cosh K_2)^{N}
G^{\rm Kln}(z,z_1,z_2),\label{kpart}
\eea
where
\be
G^{\rm Kln}(z,z_1,z_2)=\sum_{\rm closed \ polygons} z^nz_1^{n_1}z_2^{n_2}
\ee
generates all closed polygons on the $2M\times N$ lattice ${\cal L}$ with 
edge weights $z=\tanh K, z_1=\tanh K_1$, and $z_2=\tanh K_2$.  The desired 
partition function is given by
\be
Z^{\rm Kln}_{2M,N}(K,K,K)=2^{2MN}(\cosh K)^{4MN}
G^{\rm Kln}(z,z,z). \label{kpart1}
\ee

Again, it is necessary to first write $G^{\rm Kln}(z,z_1,z_2)$ 
as a multinomial in $z, z_1, z_2$ in the form of
\be
G^{\rm Kln}(z,z_1,z_2) = \sum_{m,n=0}^N T_{m,n}(z) z_1^mz_2^n, \label{genk}
\ee
where $T_{m,n}(z) $ are polynomials in $z$ with strictly positive coefficients.

The evaluation of $G^{\rm Kln}(z,z_1,z_2)$  parallels  
that of $G(z,z_1)$ for the \Mb strip.  One first maps the 
lattice ${\cal L}$ into a dimer lattice ${\cal L}_D$ by 
expanding each site into a city of 4 sites as 
shown in Fig. 2.  Orient all $K$ and $K_1$ edges 
of ${\cal L}_D$ as shown, and orient all $K_2$ edges
in the same (downward) direction as the $K_1$ edges.  
Then this defines an $8MN \times 8MN$ antisymmetric 
matrix obtained  by adding an extra term
to $A(z,z_1)$ given by  (\ref{mobpf}), namely,
\be
A^{\rm Kln}(z,z_1,z_2) = A(z,z_1) + b(z_2) \otimes G_M' \otimes H_{2N}.
\label{kleinpf}
\ee
Here, 
\bea
b(z_2) &=& \pmatrix{0&0&0&0\cr
0&0&0&0\cr 0&0&0&0\cr 0&0&0&-z_2\cr}, \nonumber \\
G_M'
&=&\pmatrix {1 & 0  &\cdots & 0 & 0 \cr
                  0 & 0  &\cdots & 0& 0 \cr
                  \vdots  & \vdots & \ddots & \vdots& \vdots \cr
                  0 & 0  &\cdots & 0 & 0 \cr
0 & 0   &\cdots & 0 & 0}.
\eea
Then, in place of   theorem ({\ref{theorem}), we now have
\be
{\rm Pf}A^{\rm Kln}(z,z_1,z_2) = \sum_{m,n=0}^N 
\epsilon_m \epsilon_n T_{m,n}(z) z_1^mz_2^n, 
\label{genk1}
\ee
from which one obtains in  a similar manner the result
\bea
&&G^{\rm Kln}(z,z_1,z_2) = \nonumber \\
&&\quad {1\over 2}\Big[{\rm Pf}A^{\rm Kln}(z,iz_1,-iz_2)
+{\rm Pf}A^{\rm Kln}(z,-iz_1,iz_2)\nonumber \\
&&\quad -i{\rm Pf}A^{\rm Kln}(z,iz_1,iz_2)
+i{\rm Pf}A^{\rm Kln}(z,-iz_1,-iz_2)\Big]. \label{klgen}
\eea

To evaluate the Pfaffian (\ref{genk1}), we note that
 the matrix (\ref{kleinpf}) can again be diagonalized in the $\{2N\}$ subspace, yielding
\bea
{\rm Pf} A^{\rm Kln}(z,z_1,z_2)&=& \sqrt {{\rm det}  A^{\rm Kln}(z,z_1,z_2)}  \nonumber \\
   &=& \prod _{n=1}^{2N}\sqrt{{\rm det} A_M^{\rm Kln}(z,z_1,z_2;\p_n)  }, \label{kleinpf1}
\eea
where 
\bea
A_M^{\rm Kln}(z,z_1,z_2;\p_n) &=& A_M (z,z_1;\p_n) \nonumber \\
&&\quad + i(-1)^{n+1}b(z_2)\otimes G_M'.
 \label{kpf}
\eea
Now we expand ${\rm det} A_M^{\rm Kln}$ in $z_2$. Since 
setting $z_2=0$ the determinant is precisely $B_M$ and
 the term linear in $z_2$, the $\{4,4\}$ element of the determinant, is by
definition $D_M$, one obtains  
\be
{\rm det} A_M^{\rm Kln}(z,z_1,z_2;\p_n)
 =B_{M}+i(-1)^nz_2D_{M}, \quad M\geq 2,
\ee
where $B_M$ and $D_M$ have been computed  in (\ref{solution}).
 This  leads to
\bea
&&{\rm Pf}A^{\rm Kln}(z,z_1,z_2)=\Big(1+{{z_1z_2}\over {z^{2}}}\Big)^N
\Big[z(1-z^2)\Big]^{MN}\nonumber \\
&&\hskip 1cm \times \prod_{n=1}^{N}
\Bigg[{\sinh (M+1)t-c(z,z_1,z_2)\sinh Mt\over \sinh t}\Bigg] \label{p}
\eea
where
\bea
c(z,z_1,z_2)&=&{1\over z(1-z^2)(z^2+z_1z_2)}\Big[
(1+z^2)(z^4+z_1z_2)\nonumber \\
&&\hskip 1.5cm +2z(z^4-z_1z_2)\cos\p_n\nonumber \\
&&\hskip 1.5cm +2(-1)^n(z_1+z_2)z^3\sin\p_n\Big].  \label{klgen1}
\eea
Setting $z_1=z_2=z$ in (\ref{klgen}) and using (\ref{p}), we obtain 
after some algebra
\bea
G^{\rm Kln}(z,z,z)&=&\Big[z(1-z^2)\Big]^{MN}\Bigg[
\prod_{n=1}^{N}2\cosh Mt(\p_n)\nonumber \\
&& +{\rm Im} \prod_{n=1}^{N}\Bigg(
{\sinh Mt(\p_n)\over \sinh t(\p_n)}D(\p_n)\Bigg)\Bigg], \label{gk}
\eea
where 
\bea
D(\p_n)
&=&{1\over z(1-z^2)}\Big[(1-z^4)-2z(1+z^2)\cos\p_n \nonumber \\
&&\hskip 2cm -4i(-1)^nz^2\sin\p_n\Big]
\eea
and Im denotes the imaginary part. The substitution of (\ref{gk}) 
into (\ref{kpart1})  now completes the evaluation
of the partition function for a $2M\times N$ Klein bottle.

For the $2\times 2$ Klein bottle, for example, we find
\bea
{\rm Pf} A^{\rm Kln}(z, z_1, z_2) &=& 1+z^4+4(z_1+z_2)z^2
-2(z_1^2+z_2^2)z^2\nonumber \\
&&+2z_1z_2(1+z^2)^2-4z_1z_2(z_1+z_2)z^2\nonumber \\
&&+z_1^2z_2^2(1+z^4),\nonumber \\
         G^{\rm Kln}(z, z_1, z_2) &=& 1+z^4+4(z_1+z_2)z^2
+2(z_1^2+z_2^2)z^2\nonumber \\
&&+2z_1z_2(1+z^2)^2+4z_1z_2(z_1+z_2)z^2\nonumber \\
&&+z_1^2z_2^2(1+z^4),
\eea
which can be verified by explicit enumerations.

For a $(2M-1)\times N$ Klein bottle we can proceed as before by first 
considering a $2M\times N$ Klein bottle with interactions $K, K_1, K_2$ 
and, within the center
two rows, interactions $K_0=K/2$ as shown in Fig. 3.  This is followed by taking 
$K_1\to \infty$ and $K_2=K$.
Thus, in place of (\ref{kpart}), we have
 \bea
Z^{\rm Kln}_{2M-1,N}(K)&=&2^{(2M-1)N}(\cosh K)^{(4M-3)N}\nonumber \\
&&\times \cosh^{2N} (K/2)G^{\rm Kln}(z,z_0,1,z),\label{part2}
\eea
where $z_0=\tanh (K/2)$ and $G(z,z_0,z_1,z_2)$ generates polygonal
configurations on the $2M\times N$ lattice. Then, as in the above, we find 
\bea
&&G^{\rm Kln}(z,z_0,z_1,z_2)=\nonumber \\
&&{1\over 2}\Big[{\rm Pf}A^{\rm Kln}(z,z_0,iz_1,-iz_2)
+{\rm Pf}A^{\rm Kln}(z,z_0,-iz_1,iz_2)\nonumber \\
&&-i{\rm Pf}A^{\rm Kln}(z,z_0,iz_1,iz_2)
+i{\rm Pf}A^{\rm Kln}(z,z_0,-iz_1,-iz_2)\Big], \label{klgen2}
\nonumber \\
\eea
where  ${\rm Pf}A^{\rm Kln}(z,z_0,z_1,z_2)$ 
is found to be given by the right-hand side of (\ref{pfa}), but with
\bea
c_1&=&(1+z_0^{2})(1-z_1z_2)-2z_0(1+z_1z_2)\cos \p_n \nonumber \\
&&-2(-1)^n(z_1+z_2)z_0\sin\p_n,\nonumber \\
c_2&=&{1\over z(1-z^2)}\Bigg\{(z^2+z_1z_2)
\Big[(1-z_0z)^2+(z-z_0)^2\Big]\nonumber \\
&&\hskip 1.5cm +2(z-z_0)(1-z_0z)\Big[(z^2-z_1z_2)\cos\p_n\nonumber \\
&&\hskip 1.5cm +(-1)^n(z^2z_1+z_2)\sin\p_n\Big]\Bigg\}, \label{c1c2}
\eea
expressions which are valid for arbitrary $z, z_0, z_1,$ and $z_2$.  
For $z_0=\tanh (K/2)$, the case we are considering, (\ref{c1c2}) reduces to
\bea
c_1&=&{2z_0\over z}\Big[
1-z_1z_2-z(1+z_1z_2)\cos \p_n\nonumber \\
&&\hskip .8cm -(-1)^nz(z_1+z_2)\sin\p_n\Big],\nonumber \\
c_2&=& {2z_0\over z^2}\Big[z^2+z_1z_2+  z(z^2-z_1z_2)\cos\p_n\nonumber \\
&&\hskip .8cm +(-1)^nz(z^2z_1+z_2)\sin\p_n\Big],
\label{c1c2a}
\eea
which   reduces further to (\ref{cc}) after setting $z_2=0$.
The explicit expression for the partition function is now obtained
by substituting (\ref{klgen2}) into (\ref{part2}).

\section{The bulk limit and finite-size corrections}
 In the thermodynamic limit, our solutions of the Ising partition
function  give rise to a bulk free energy
\be
f_{\rm bulk}(K) = \lim_{M,N \to \infty} {1\over {2MN}}\ln Z(K) \label{bulk}
\ee
identical to that of the Onsager solution \cite{onsager}.
Here, $Z(K)$ is any one of the 4 partition functions.
Indeed, using the solution (\ref{partmob}) for the $2M\times N$
\Mb strip, for example, one obtains
\bea
f_{\rm bulk}(K) &= & C(K) + \lim_{N\to \infty}(2N)^{-1} \sum_{n=1}^N t(\p_n) \nonumber \\
  &=& C(K)  +{1\over {2\pi}} \int_0^\pi d\p\  t(\p) \nonumber \\
  &=& C(K)  +{1\over {2\pi^2}} \int_0^\pi d\p 
      \int_0^\pi d\theta  \nonumber \\
&& \ln\Big[2\cosh 2K \coth 2K -2(\cos\theta +\cos\p)\Big],
  \label{bulk1}
\eea
where $C(K) =[ \ln (2\sinh 2 K)]/2$.  
This expression is the Onsager solution (steps
leading to the last line of (\ref{bulk1}) can be found in \cite{huang}).  
It is well-known that $f_{\rm bulk}(K)$ is  
singular at the critical point 
$\sinh 2K_c=1$ or $2K_c=\ln(\sqrt 2 +1)$.

For large $M$ and $N$, one can use
 the Euler-MacLaurin summation formula to evaluate corrections to the bulk free energy,
an  analysis first  carried out by Ferdinand and Fisher \cite{ff}
for the Kaufman solution of the Ising model on a torus.   
Generally, for large $M$ and $N$, 
 we expect to have
\bea
\ln  Z_{2M,N}(K)& =& 2MN f_{\rm bulk}(K) + N c_1(\xi,K) \nonumber \\
&&+ 2Mc_2(\xi,K) +c_3(\xi,K) + \cdots
\eea
where $\xi = N/2M$ is the aspect ratio of the lattice. 
For the purpose of comparing with the conformal field theory \cite{cardy},
it is of particular interest to analyze corrections at the critical point.
Following  \cite{ff} as well as similar analyses for  dimer 
systems \cite{luwu99,ferdinand},  
we have carried out such analyses for our solutions as well 
as for the solution of the
Ising model under cylindrical boundary conditions \cite{onsager,mccoywu}. 

For the $2M\times N$ \Mb strip, for example, one starts with 
the explicit expression (\ref{partmob}) 
of the partition function, and uses the Euler-MacLaurin 
formula to evaluate corrections to the bulk free energy.
The analysis is lengthy, even at the critical point $K_c$.
We shall give details elsewhere and quote here only the results:
\bea
c_1(\xi,K_c) &\equiv& c_1^{\rm Mob} =I - K_c = -0.087\ 618\ ..., \nonumber \\
c_2(\xi,K_c) &=& 0, \nonumber \\
c_3(\xi,K_c) &=& -{1\over 2}\ln 2
 +{1\over 12}\ln\Bigg[{2\vt_3^2(0|i\xi)\over \vt_2(0|i\xi)
\vt_4(0|i\xi)}\Bigg]\nonumber \\
&&+{1\over 2}\ln\Bigg[1+{\vt_3(0|i\xi/2)-\vt_4(0|i\xi/2)\over {2}\vt_3(0|i\xi)}\Bigg]
\eea
where
\begin{eqnarray*}
I&=&{1\over {2\pi}}\int_0^{\pi}\ln\Big(\sqrt{2}\sin\p+ 
\sqrt{1+\sin^2\p}\Big)d\p  \\
&=&0.353\ 068\ ... 
\end{eqnarray*}
and $\vt_i(u|\tau), i=2,3,4$, are the Jacobi theta functions  \cite{gr}
\bea
\vt_2(u|\tau)&=& 2\sum_{n=1}^{\infty}q^{(n-{1\over 2})^2}\cos (2n-1)u,
\nonumber \\
\vt_3(u|\tau)&=&1+2\sum_{n=1}^{\infty}q^{n^2}\cos 2nu,  \nonumber \\
\vt_4(u|\tau)&=& 1+2\sum_{n=1}^{\infty}(-1)^n q^{n^2}\cos 2nu, \label{thetafunction}
\eea
with $q=e^{i\pi \tau}$. 
 For the $2M\times N$ Klein bottle, we find
\bea
c_1(\xi,K_c) &=& 0 \nonumber \\
c_2(\xi,K_c) &=& 0 \nonumber \\
c_3(\xi,K_c) &=&  {1\over 6}\ln\Bigg[{2\vt_3^2(0|2i\xi)\over 
\vt_2(0|2i\xi)\vt_4(0|2i\xi)}\Bigg]\nonumber \\
&&+\ln\Bigg[1+\sqrt{\vt_2(0|2i\xi)\over 2\vt_3(0|2i\xi)}\Bigg].
\eea

If one takes the limit of $N\to \infty$ ($M\to \infty$) first while keeping
$M$ ($N$) finite, one obtains 
\bea
\lim_{N\to\infty} {1\over {N}}\ln Z_{2M,N}(K_c)
&=& 2M f_{\rm bulk}(K_c) + c_1 \nonumber \\
&&+{ {\Delta_1}/ {2M}}
+O(1/M^2),\label{Nlimit} \nonumber \\
\lim_{M\to\infty} {1\over {2M}}\ln Z_{2M,N}(K_c)
&=& N f_{\rm bulk} (K_c) +  c_2 \nonumber \\
&&+{ \Delta_2/ {N}}
+O(1/N^2), \label{Mlimit}
\eea
where $c_1,c_2,\Delta_1,\Delta_2$ are constants.
We list our resluts in the  table below.  Also listed are  
values for the Ising model with toroidal boundary conditions 
taken from \cite{ff}, and values for the 
cylindrical boundary conditions computed using the solution of 
\cite{mccoywu}. Our values of $\Delta_1$ and $\Delta_2$ imply 
a central charge
\be
c = 1/2
\ee
for the Ising model. This is
consistent with the conformal field theory prediction \cite{cardy}.
\begin{center} 
\begin{tabular}{cccccc} \\
\hline\hline
 & M\"obius &Klein &Cylindrical & Toroidal \\
\hline
$c_1$ & $c_1^{\rm Mob}$& 0& $c_1^{\rm Mob}$&0 \\
$c_2$ &0  &0 & 0 &0 \\
$\Delta_1$ & $ \pi /48$ & $ \pi /12 $& 
 $\pi /48$ & $ \pi /12 $ \\
$\Delta_2 $&  $\pi /48 $ & $ \pi  
/48$  &  $\pi /12$ & $ \pi/12$\\
\hline\hline
\end{tabular}
\end{center}

\section{Summary}
We have solved and obtained  close-form expressions for the partition 
function of an Ising model on finite \Mb strips and Klein bottles with 
a uniform interaction $K$. The solution assumes different forms 
depending on whether the width of the lattice is even or odd.
For a $2M\times N$ \Mb strip, where $2M$ is its width, the partition 
function $Z^{\rm Mob}_{2M,N}(K)$  is given by (\ref{partmob}) 
with $F_\pm$ given by (\ref{fpm}).  For a $(2M-1)\times N$ \Mb 
strip we employ a trick by first considering a $2M\times N$ lattice
and then ``fusing" it into the desired lattice by coalescing
two rows of spins.  The  resulting
partition function $Z^{\rm Mob}_{2M-1,N}(K)$ is given by (\ref{part1})
in which the generating function $G(z,z_0,z_1)$ is 
(\ref{oddgen2}) with the Pfaffians given by (\ref{pfa}).

For a $2M\times N$ Klein bottle the partition function 
$Z^{\rm Kln}_{2M-1,N}(K)$
is given by (\ref{kpart1}) in which the generating
function $G^{\rm Kln}(z,z,z)$ is given by (\ref{gk}).
For a $(2M-1)\times N$ Klein bottle the partition function 
$Z^{\rm Kln}_{2M-1,N}(K)$ is given by 
 (\ref{part2}) in which the generating function 
$G^{\rm Kln}(z,z_0,1,z)$, $z_0=\tanh (K/2)$, is computed
 using (\ref{klgen2}).  All solutions yield the same Onsager bulk
free energy (\ref{bulk1}).

We have also presented results of finite-size analyses of our 
solutions as well as those of the Ising model under cylindrical 
and toroidal boundary conditions at criticality.  
The analyses yield a central charge $c=1/2$ in agreement with the
conformal field prediction \cite{cardy}.

\section*{Acknowledgement}
Work has been supported in part by National Science Foundation Grant
DMR-9980440.

\end{document}